# CHRONIC DISEASES AND BIORHYTHM'S PHASE-SHIFT DYNAMIC. LYAPOUNOV CRITERIA FOR HEALTH STABILITY

by Prof. Maria Kuman, PhD

Health and Happiness Books, 1414 Barcelona Dr., Knoxville, TN 37923, USA

holisticare1@gmail.com

Abstract

When stressors are present, they cause delays (phase shifts) in the biorhythms of the body because the body needs to stop its normal habitual work and mobilize for respond to the stressors. The genetically inherited weak organ, being the organ with lowest energy in the body, will not have enough energy to propel the phase shifts, and they will accumulate there; this will make its functioning sluggish and its biorhythms weakly integrated. When the accumulated phase shifts surpass the ultimate level, which the genetically weak organ can tolerate, its biorhythms become desynchronized, which will lead to chronic disease of this organ.

This article discusses the dynamic of phase shift changes caused by stress and uses Lyapounov's criteria for stability to evaluate how close the genetically weak organ of an individual is to becoming chronically sick. Both, the effect of self-calming for the organ's biorhythms and the effect of irregular self-calming for the body's biorhythms are considered.

Words: 156

1.  **Introduction**

We don't understand the nature of chronic pain and chronic diseases and we don't know how to treat them to achieve cure. Our contemporary medicine can only offer pain-killing drugs for temporary relief of chronic pain.

In a number of books, the author (M. Kuman, 1993, 1995, 1997) has offered a mechanism of onset of chronic diseases by strong or prolonged stress. (Prolonged stress is a series of stressors acting through relatively small time intervals, which allows accumulation of their effect on the body). 'Chronic' means 'slow'. The author has explained that not only the onset of each chronic disease is slow; by using the non-equilibrium theory of Prigogine, she has explained why the cure of chronic diseases is slow, difficult and unpredictable.

Considering this, we should be looking for prevention rather than cure of chronic diseases. Prevention would require finding the changes in the body that precede the chronic disease and finding a way to eliminate these changes and prevent the disease.

This article offers a mathematical model, which allows prediction of the onset of chronic diseases. Strong or prolonged stress causes large delays because the body needs to stop its due work and mobilize for response to the stressors. The periodic reactions in the body are called biorhythms and their delays are called phase shifts (G. G. Luce, 1970, p. 14, 9).

Strong or prolonged stress creates large phase shifts, for which the body cannot compensate. These large phase shifts will accumulate in the genetically inherited weak organ, whose low energy is not enough to propel them.

The process of accumulation is slow and accompanied by "moving pain", which the doctors presently consider "psychosomatic". This means the patients just imagine they have pain moving from organ to organ. It is always ignored because we do not have equipment sensitive enough to detect the subtle changes related to these symptoms.

I think the "moving pain" reflects the movement of large delays (phase shifts) caused by strong or prolonged stress, which follow the pathway of organ's maximal activity. The delays (phase shift) circulate until they reach the genetically inherited weak organ, whose weakly integrated biorhythms will absorb them. The accumulated phase shifts (delays) will further disintegrate the biorhythms of the weak organ, and they will become more desynchronized.

When the ultimate stress $E_c$ is reached, which the weak organ can endure, its biorhythms will become completely desynchronized. *"When biorhythms are desynchronized this is a step toward a disease."* (See G. G. Luce, 1970, p. 42, see also M. C. Moore-Ede et al. 1983). This will lead to dysfunction, or the genetically inherited weak organ will suffer functional (chronic) disease. (For the relationship of chronic diseases to biorhythm de-synchronization, see also H. Selye, 1974).

With aging the stress of life will more and more desynchronize the biorhythms of the genetically weak organ (see G. G. Luce, 1970, p. 43). This explains why chronic diseases are more frequent among the elderly. Cancer is also more frequent at advanced age. *"Altered phase relations among biorhythms await study in health and disease, ranging from central nervous system disease to cancer"* (F. Halberg, 1960).

Hence, the phase shifts changes and their dynamic could and should be used as basic characteristics of health (synchronized biorhythms) and disease (desynchronized biorhythms). Phase shifts, measured before the onset of chronic disease or cancer (see G. G. Luce, 1970, p. 68), can be used for very early diagnosis of these diseases.

Phase shifts have been used in science as basic characteristics of the stability of systems: in superconductivity, laser emission, and electric power systems. *Here phase shifts will be used as basic characteristics of the functional stability of organs in the body. In living beings, phase shifts are delays caused by stress and their monitoring will allow us to evaluate the harm caused by strong or prolonged stress*.

This article uses a nonlinear mathematical model to describe the dynamic of biorhythm changes caused by stress. The biorhythms are considered a system of *n* nonlinearly related oscillators. The model offered here is a generalization of the three dimensional model used by the author to explain the integration of the electrical signals of vision, hearing, and vestibular apparatus (M. Kuman(ova), 1984).

The three signals were considered as three nonlinearly related oscillators. With this model, we were able to explain the peculiar changes in vision and hearing, experimentally observed after vestibular stimulation of patients with vestibular problems. The changes in hearing intensity, and the shape and intensity of images, were successfully explained as phase shift changes cased by the vestibular stimulation with cold water.

This makes us believe that the generalized model offered here, will be just as useful in explaining the health stability of a body under stress as our three-dimensional model was in explaining the integration of visual, hearing, and vestibular signals. Modifications of the generalized model offered here can also be used to describe other systems of biorhythms existing at different levels (body, organ, or cell).

The mathematical model described in the next paragraph will not only help us understand how stress causes chronic diseases and cancer, it will help us learn how to prevent these diseases.

## 2. Nonlinear Mathematical Model

*"The organization of the biorhythms seems to be plastic rather than rigid." (F. Halberg, 1960, p. 289).* The biorhythm organizations at each level: cell, organ, or body, show unlimited flexibility and plasticity in their abilities to adapt to any changes in the environment. This unlimited plasticity could be explained only with nonlinear relations among the biorhythms.

Illustration of nonlinear relations among the biorhythms of two human bodies are the observed: entrainment of the sleeping pattern of the husband by the sleeping pattern of the wife and vice versa (A. Winfree, 1987, p. 113, 120); entrainment of the biorhythms of a blind child by the circadian biorhythms of the mother (G. G. Luce, 1970, p. 37), etc. Such mutual entrainment is specific only for nonlinear systems (R. Minorski, 1964, p. 458).

Hence, the biorhythms in the body must be nonlinearly related and only nonlinear equations can describe their dynamical plasticity and flexibility at any level - cell, organ, or body. Taking this into consideration, the following nonlinear differential equation is offered for description of the dynamic of phase shift changes:

$$J\omega_i' + D_i\omega_i + \Sigma d_{ij}(\omega_i - \omega_j) + \Sigma K_{ij}\cos\hat{I}_{,ij} = \Sigma B_{ij} = B_i \qquad (2.1)$$

Where $\omega_i' = d\omega_i/dt$ ($\omega_i = d\theta_i/dt$) characterize the dynamic of phase shift changes; $J_i$ is inertial coefficient. (Simplified version of equation (1) is the adaptive frequency model of B. Ermentrout of 1991)

The second term in eqn. (1) represents the elastic force. The greater is the phase deviation from norm $\omega_i = d\theta_i/dt$, the greater is the force trying to bring the system back to its initial state. The proportionality constant $D_i < 0$ is called coefficient of selfcalming.

The third term reflects the calming effect of other biorhythms j on the considered biorhythm i. The calming force trying to return the system to its initial state will be proportional to the difference in phase shift changes of the considered biorhythm i and the functionally related to it biorhythm j. The coefficient of proportionality $d_{ij}$ is called coefficient of mutual effect of calming.

The fourth term reflects the biorhythm interaction. $\theta_{ij} = \theta_i - \theta_j$, which is the mutual phase shift of any couple of interrelated biorhythms. $K_{ij}$ is a nonlinear coefficient.

The term $B_{ij}$ is the energy of integration of the considered biorhythm i with other biorhythms j (j = 1, 2, 3,...,n). $B_{ij}$ will be different for different organs even within one individual. It will vary from individual to individual and will decrease slowly with aging. Since the decrease is slow, as a first approximation, for a limited interval of time, $B_{ij}$ could be considered a constant.

The method of Lyapounov's functions (V. M. Matrosov and C. H. Vassileva, 1964); I. G. Malkin, 1952, pp. 37, 79; A. Ghizzetti, 1966, p. 19-107) is a useful tool for investigation of dynamic stability. Lyapounov criteria are used here to characterize the health stability and evaluate the chances to become chronically sick.

Following the Lyapounov's method, we can write the equation of the boundary curve S of the region of dynamic stability G of a system of n nonlinearly related equations

$$S(G): V(0, \omega) - V_L = 0 \tag{2.2}$$

where

$V_L = V(-\theta_s, 0)$; index s means stable state.

On Fig. 1, the boundary curves and the region of dynamic stability (shaded area) are given in two-dimensional phase space $\omega/\theta$, for the simplest case - a system of two nonlinearly related biorhythms. The boundary of the real region of dynamic stability might be determined from the initial conditions at different limited perturbations.

As one can see in the figure, the stability trajectory is a distorted ellipse, because the "force" or the "spring" that tries to bring the system to equilibrium is nonlinear. This makes the coefficients $D_i$ and $d_{ij}$ nonlinear. The trajectory does not surround the origin, but a simple change of variable can move the origin to $\theta = \theta_2$ ($\theta_2$ is a stable equilibrium point).

Now, at zero displacement the maximum of the dynamic of phase shifts is at the positive value $+\omega_2$ and its minimum at the negative value $-\omega_2$. The displacement from equilibrium has a maximum positive value $\theta_{max}$ or negative value $\theta_0$ when $\omega = d\theta/dt$ is zero.

The behavior of this system, when stable, is thus similar to the spring mass system. When the phase shift $\theta > \theta_{max}$, the "spring" reverses its force, and the phase shift deviation from the norm increases rather than decreases. Then the system loses synchronization.

This was in the case of two nonlinearly related oscillators, the description of which require two-dimensional phase space. In the common case - a system of *n* nonlinearly related oscillators - the region of dynamic stability G and its boundary hypersurface S are in *n*-dimensional phase space. To define them, we need the concrete form of the Lyapounov's function V (see eq. (2.2)).

Different methods are developed for finding the region of dynamical stability G (Williams, 1973) of a system of <u>n</u> nonlinearly related equations (see for this purpose V. M. Matrossov and C. H. Vassileva, 1984). We consider the Lour'e matrix method as most convenient for our case (V. M. Matrossov and C. H. Vassileva, 1984; J. L. Willems, 1970).

In order to use the Lour'e method, we have to write our eqn.(1) in a matrix form. We can write (2.1) in Lour'e form as

$$X = A X + B F(X). \qquad (2.3)$$

In the Lour'e form, the nonlinear coefficients $D_i$, $d_{ij}$, $K_{ij}$ are separated as a term BF(X). This nonlinear term will describe the flexibility and plasticity of a biorhythm system in its reaction to external influences.

$$\begin{vmatrix} J^{-1}D & 0 \\ I & 0 \end{vmatrix} \quad \begin{vmatrix} J^{-1} & T^{-1} \\ 0 & 0 \end{vmatrix} \quad C = \begin{vmatrix} 0 & T \end{vmatrix}$$

$$X^T = | \omega \quad \Delta\theta_{abs} |; \qquad \omega = | \omega_1, \omega_2,...,\omega_n |^T; \qquad (2.4)$$

$$\Delta\theta_{abs} = \theta_{abs} - \theta_{abs}^s; \qquad \theta = T \Delta\theta_{abs};$$

T - bond matrix with two nonzero elements in each row +1 and -1.

F(θ) - vector function with elements $F_i$ with only one argument i. $\theta_{abs}^s$ - the absolute phase shift, leaving the system stable.

$$J = \text{diag}[(J_i)]; \quad D = \text{diag}[(-D_i)]; \quad I = \text{diag}[(1)]; \quad 0 = [(0)];$$

The V - function in this case is given by the expression

$$V_1(X) = X^T P X + 2 \int F^T(y) Q \, dy \tag{2.5}$$

The region, where the function V is positively determined is given by

$$\theta_i F(\theta_i) > 0; \quad i = 1, 2, \ldots, m = n(n-1)/2$$

The derivative of V, V(X), is negatively determined at

$$Q = I; \quad \begin{vmatrix} J + b\, JIJ & -c\, DIJ \\ -c\, DIJ & c\, DID \end{vmatrix}$$

b, c - arbitrary scalar coefficients

$$0 > b > b_0 = -1/J_0;$$

If V is positively determined and its first time derivative negatively determined it can be considered as Lyapounov's function. At D = 0 (absence of selfcalming effect)

$$V_1(D=0) = \omega^T J \omega + b\, \omega^T JIJ \omega + 2 \int F^T(y) \, dy \tag{2.6}$$

At D = 0 the quadratic form $X^T P X$ is positive only at b = 0. But b = 0 means that the irregularity of the absolute selfcalming is excluded from consideration or that the selfcalming is regular. In reality, the selfcalming of each biorhythm is not a regular process, since the continuous influence of stressors interferes with the selfcalming process (G. G. Luce, 1970, p. 138).

To give an account of the irregularity of the selfcalming process, we have to add to expression (2.5) new terms

$$V_2 = V_1 + a\, \omega^T D \omega + 2 \int F^T(Tz)\, TDJ^{-1} dz \tag{2.7}$$

The new terms at a < 0 are positively determined and their derivatives are negatively determined. Hence $V_2(X)$ is a Lyapounov's function and can be used for investigation of the dynamical stability of a system of biorhythms with irregular selfcalming.

In order to get the new V-function, strongly following the Lour'e method, we must change the content of P and Q

$$Q \rightarrow Q_D = (I + a D J^{-1})$$

$$P \rightarrow P_D = (P + F^T DF)$$

where $F = [I\ 0]$; $\omega = F X$;

Experimental evidence indicates that the phase shifts are tissue specific and since the organs are tissue specific, different organ's tissues need different time to adapt to a new (reverse) light schedule: *"In liver parenchyma in mice an inversion of mitotic rhythms was detected 9 days after lightening reversal, while the mitotic rhythm in pineal epidermis had not yet been fully reversed"* (F. Halberg, 1960, p. 301).

Since each healthy organ is a homogeneous media consisting of one type of tissue-specific cells (liver, heart, kidneys, lungs, etc.), it seems plausible to admit that each organ will have only one type of phase shift changes, and irregular mutual intercalming effect should not be expected (which is specific for inhomogeneous media). If this is true*, the biorhythm stability of an organ would be determined by $V_2$.*

It was shown experimentally (G. G. Luce, 1970) that in the body the mutual intercalming effect of the biorhythms is irregular. For example, after more than 6-hour flight east or west, all the biorhythms must be shifted to adapt to the new local time. The adaptation time is different for different biorhythms - it is 6 days for the temperature rhythm, 8 days for the cardiac rhythm, etc.

If so, if we want to use the last expression (2.7) for body biorhythms, eqn. (2.7) should be modified with additional term reflecting the irregularity of the intercalming effect. *The Lyapounov's function for body rhythms is:*

$$V_3 = V_2 + e \, \omega^T \, T^T \, D \, T \, \omega + 2 \int F(Tz) \, TT^T DTT^{-1} dz \tag{2.8}$$

where $D = \text{diag}[(d_{ij} + d_{ji});\ i = 1, 2, ..., n-1;\ j = 2, 3, ..., n;\ (i<j);\ e$ - arbitrary scalar coefficient.

The new terms are positively determined and their derivatives negatively determined, thus the last expression might be used as Lyapounov's function and will determine the dynamical stability of body system of biorhythms with irregular calming and intercalming effects.

In order to get the new V-function, strongly following the Lour'e method, the content of P, Q, D must be changed in the following way:

$$P_D^{ir} = P_D + e \, F^T D F;$$

$$Q_D^{ir} = Q_D + e \, D \, J^{-1} = I + a \, D \, J^{-1} + c \, D \, J^{-1};$$

$$D \to (D_D^{ir})_{ii} = - D_{ii} - \Sigma \, d_{ij}; \quad (D_D^{ir})_{ij} = d_{ij};$$

$$D \to \Sigma(d_{ij} + d_{ji}) \text{ at } i = j; \quad D \to - d_{ij} - d_{ji};$$

*Thus, the Lyapounov's function $V_3$ describes the biorhythms of a body under stress with irregular calming and intercalming biorhythms. The hyper-surface S determines the region of dynamical stability of the body.*

The brain tissue is inhomogeneous because conducting ganglia and neurons are imbedded in the gray brain tissue, which is semiconductor. Thus, the biorhythms in the brain should be with irregular calming and intercalming and their stability should be described by $V_3$.

Task for the future will be to find the absolute value of the limit of phase shift changes $\Delta\theta_{abs} = \theta_{abs} - \theta_{abs}^s$, which leave the system in the region of dynamic stability for sure. This means finding the limit of stress (and the created by it phase shift changes) that allows sustaining the state of health.

Each time a strong stressor (or a series of stressors acting simultaneously or through relatively short time intervals), creates a phase shift $\theta > \theta_{max}$ in an organ or body, the system of biorhythms will lose its stability and the existing biorhythm organization will be destroyed. (Experimental evidence for existence of $\theta_{max}$ can be found in A. Winfree, 1987, p.170).

## 3. Conclusion

Thus, computer simulated geometrical representation of eqn. (2.1) could illustrate quite well how close to pathology a system of biorhythms is. Torus representation will be used in our next article, which will illustrate that when the phase shift resetting curve links through the hole of the torus, which is called torus resetting, disease can be expected.

Words: 3,068

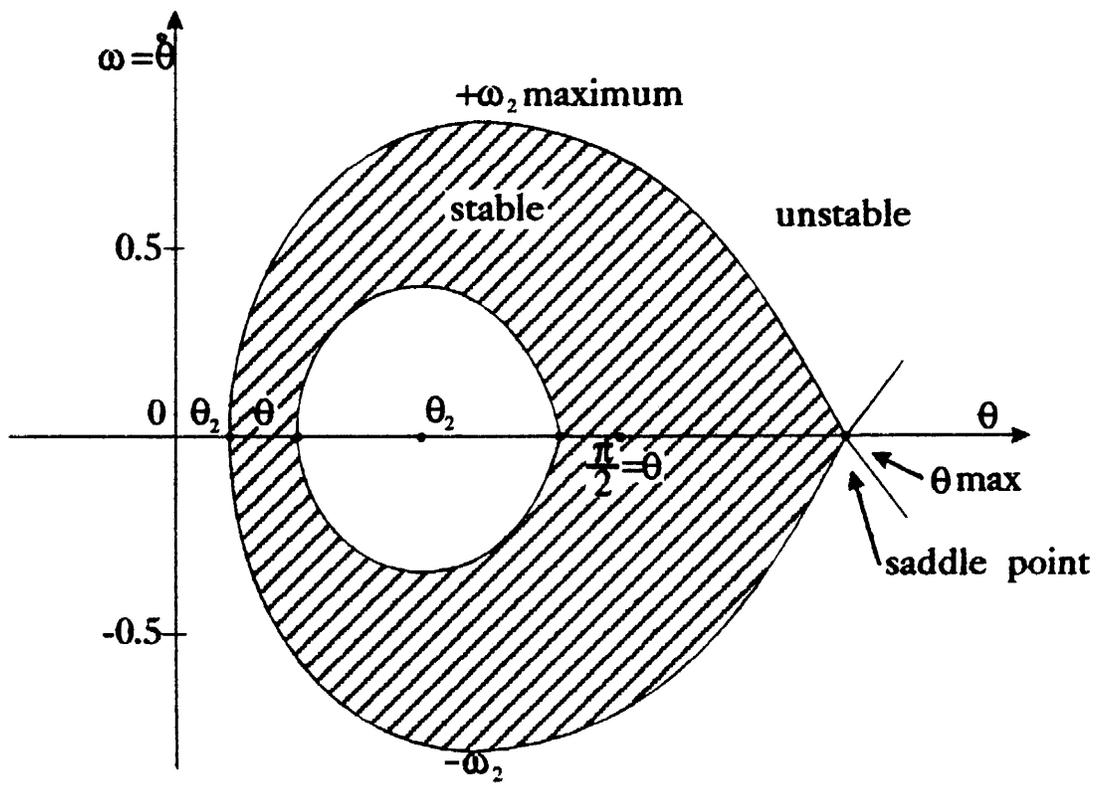

Figure 1